\documentclass[aps,pre,twocolumn,showpacs,superscriptaddress]{revtex4-1}
\usepackage{graphicx}
\usepackage{epsfig}
\usepackage[nooneline]{subfigure}
\usepackage[]{natbib}
\usepackage{amsmath}
\usepackage{float}

\begin{document}

\title{Asymptotic densities of ballistic L\'evy walks}

\author{D. Froemberg}
\affiliation{Max Planck Institute for the Physics of Complex Systems, N\"{o}thnitzer Str. 38, D-01187 Dresden, Germany}
\author{M. Schmiedeberg}
\affiliation{Insitut f\"{u}r Theoretische Physik 2: Weiche Materie, 
Heinrich-Heine-Universit\"{a}t D\"{u}sseldorf, 40204 D\"{u}sseldorf, Germany}
\author{E. Barkai}
\affiliation{Department of Physics, Institute of Nanotechnology and Advanced Materials, 
Bar-Ilan University, Ramat-Gan, 52900, Israel}
\author{V. Zaburdaev}
\affiliation{Max Planck Institute for the Physics of Complex Systems, N\"{o}thnitzer Str. 38, D-01187 Dresden, Germany}


\begin{abstract}  We propose an analytical method to determine 
the shape of density profiles in the asymptotic long time limit for a broad class of coupled continuous time random walks 
which operate in the ballistic regime. In particular, we show that different scenarios of performing a random walk step, 
via making an instantaneous jump penalized by  a proper waiting time or via moving with a constant speed, 
dramatically effect the corresponding propagators, despite the fact that the end points of the steps are identical. 
Furthermore, if the speed during each step of the random walk is itself a random variable, 
its distribution gets clearly reflected in the asymptotic density of random walkers.
These features are in contrast with more standard non-ballistic random walks.
\end{abstract}
\maketitle



\section{Introduction}

Ballistic motion is ubiquitous and often lies at the origin of stochastic transport phenomena. 
Swimming bacterial cells, scattered photons, or atoms in optical lattices 
\cite{Sagi12,Kess12,Dech14,Kor04,Matt11,Baud14,Pfei99,Bart08} 
move with fixed speeds between tumbles and collisions which reset their velocities to new values 
and eventually lead to the randomization of the dispersal process. 
However, there is a large class of diffusion processes where the ballistic behaviour 
of particles between the reorientation events is retained at larger scales \cite{Bart08,Qdot,Qdot2,plasma}.
Such ballistic diffusion processes can be modelled on the stochastic level by certain 
classes of random walks, the so-called L\'evy walks, where the displacements and the time intervals between the successive 
direction renewals are coupled \cite{ScherLax73, Bou90, MetzKlaf00, VZ14rev}.
What all these ballistic random walks do have in common is that the 
probability density function (pdf) of the time intervals has a slowly decaying power law tail
$\psi(\tau)\propto t^{-1-\gamma}$ with $0<\gamma<1$. This pdf lacks a typical scale (has a divergent mean) and gives 
rise to the ballistic scaling of the whole density profile of diffusing particles.
Depending on the implementation of the space-time coupling one has to distinguish 
between jump and velocity models \cite{ZumKlaf93}. In velocity models a particle moves at a certain random 
but constant velocity during the flight time and at each renewal a new velocity and a new flight time are chosen 
from the corresponding probability distributions. 
In the jump models \cite{ShlKlW82}, the displacement of the particle takes place instantaneously but is penalized by a certain
waiting time -- longer jumps will require longer waiting before another jump may occur.
In the wait-first model the particle jumps at the end of the waiting time, 
whereas in the jump-first model the particle jumps and then rests for the corresponding waiting time.
Such random walks were previously considered in the literature  \cite{Shle87,Be-Ke04,Jurl12,Aki14,Mag12,Liu13,Kot95,Meer06,Stra11}, 
and recently also with a method of infinite densities 
\cite{Reb14ICD} in the sub-ballistic superdiffusive regime (flight or waiting times with finite mean, $1<\gamma<2$). 
However, in general the exact analytical solutions describing the density of random walking particles are rare and 
can be obtained only for some particular values of $\gamma$ \cite{ZumKlaf93,VZ1}. 

In this paper we suggest a method to compute explicitly the asymptotic densities of random walks 
in the regime of ballistic scaling. We show that this approach can be applied both 
to coupled jump models and to the models with random velocities, 
the latter being related to weak ergodicity breaking. 
Remarkably, the choice of whether the particle jumps at the beginning or end of the waiting time (jump models), 
as well as the distribution of velocities (velocity models) has a dramatic effect on the shape of the particles' density, 
a result which could not be obtained from the standard asymptotic analysis routinely employed in random walks.

\section{Models}
\begin{figure}[t]\includegraphics[width=.44\textwidth]{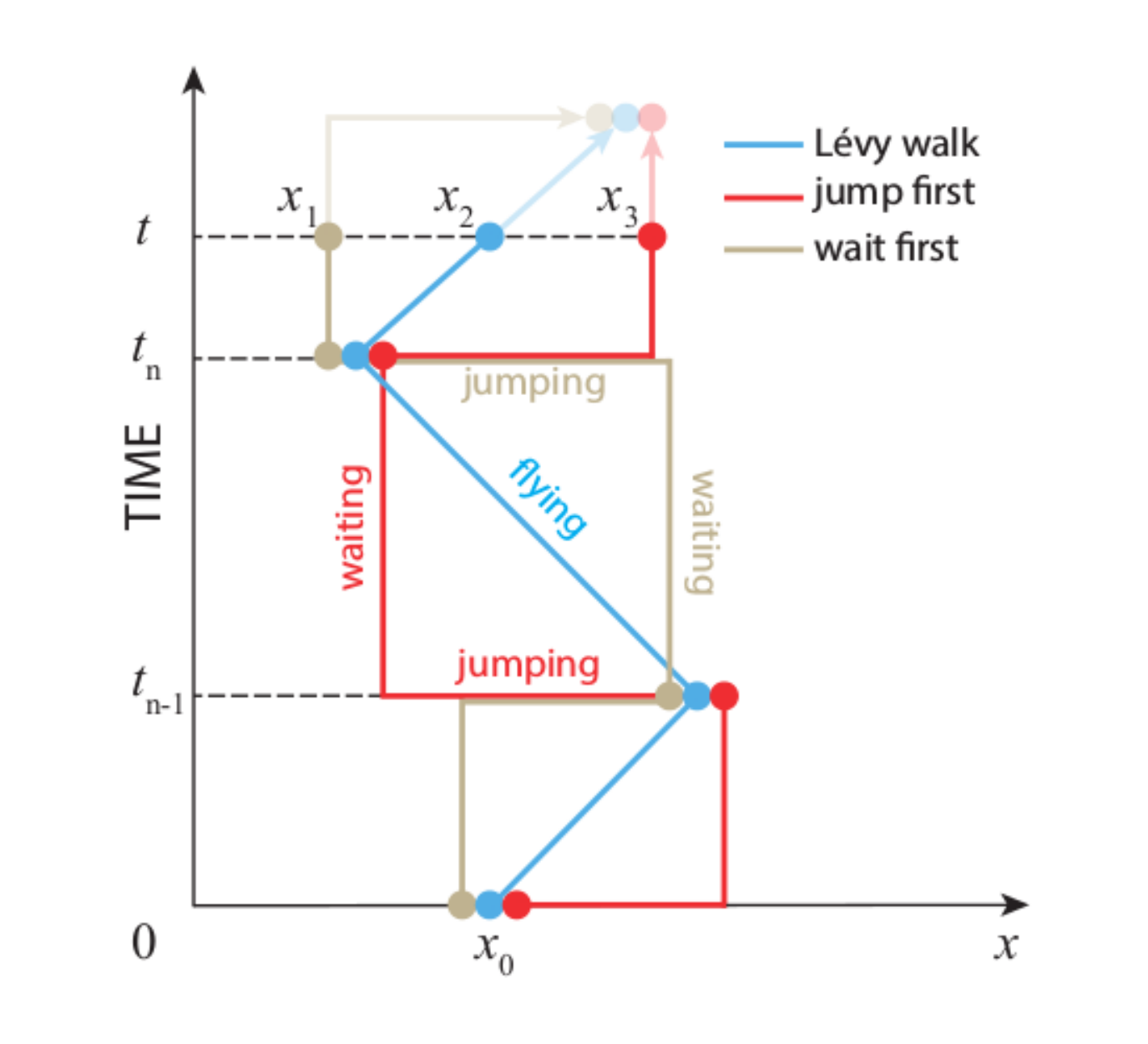}
\caption[Trajectories of the three stochastic models investigated in the text.]
{(Color online) Trajectories of the three stochastic models investigated in the text. 
L\'evy walk (blue), wait-first (green) and jump-first (red) 
models produce trajectories which take different paths but pass through the same end points on the $(x,t)$ plane. 
Therefore, only the very last, uncompleted step distinguishes these three models. $t_n$ is the last renewal event before the 
measurement time $t$, and $t-t_n$ is called the backward recurrence time}
\label{RWmodels} 
\end{figure}

Let us briefly recap the basic microscopic equations of the models considered in the present paper.
Here we restrict ourselves to the one-dimensional case.
In the jump models the jumps are penalized by a waiting time. For the jump-first model, the particle
performs a jump according to a jump length pdf $f(x)$ and then waits for a time $\tau$ drawn from
an $x$-dependent waiting time pdf $\psi(\tau|x)$. This process is then renewed. For the wait-first model, 
the particle waits first for a time $\tau$ drawn from the waiting time pdf $\psi(\tau)$ and then jumps
over a distance $x$ given by $f(x|\tau)$ at the end of the waiting time (see Fig. \ref{RWmodels}).
Thus it is easy to construct a balance equation for the probability density of particles $Q_{\alpha}(x,t)$ 
to end a step at $x$ at time $t$
($\alpha=$ ``j'' or ``w'', denoting jump- or wait-first models, respectively). 
The pdf that the step ends at $x$ at time $t$ is determined by the joint pdf $K(x^\prime,\tau)$ that the actual 
jump had length $x'$ and duration $\tau$, provided that the particle ended its previous jump 
at $x-x'$ at $t-\tau$,
\begin{eqnarray}
Q_{\alpha}(x,t)&=&\int_{-\infty}^{\infty}dx'\int_{0}^{t}K_{\alpha}(x',\tau)Q_{\alpha}(x-x',t-\tau)d\tau
\nonumber\\
&&+\delta(x)\delta(t).
\label{flux}
\end{eqnarray}
Here the coupled jump kernels for each model are $K_{\text{w}}=\psi(\tau)f(x'|\tau)$ and 
$K_{\text{j}}=f(x')\psi(\tau |x')$, and the last summand is the initial condition. 
Note that the distinction between the $Q_\alpha$ at this point is only a formal one.
The $Q_\alpha$ are mathematically the same for both jump models.
Still, both models are qualitatively different, and the $K_\alpha$ contain the information 
about which variables are dependent or independent 
(and thus implicitly on what happens between the start and end of a jump) which
comes into effect in the following equations for the densities.
We assume that the process starts at $x=0$ and $t=0$ so that the beginning of the 
first waiting time interval coincides for all sample trajectories.
The expressions for the density of particles are then:
\begin{eqnarray}
p_{\text{w}}(x,t) &=& \int_{0}^{t}\Psi(\tau)Q_{\text{w}}(x,t-\tau)d\tau 
\label{pdfw} \\
p_{\text{j}}(x,t)&=&\int_{-\infty}^{\infty}dx'\int_{0}^{t}f(x')\Psi(\tau|x')Q_{\text{j}}(x-x',t-\tau)d\tau
\nonumber\\
\label{pdfj}
\end{eqnarray}
where $\Psi(t)=1-\int_{0}^{t}\psi(\tau)d\tau$ and $\Psi(\tau|x)=1-\int_0^{\tau}\psi(\tau'|x) d\tau'$ 
are the survival probabilities, which are equivalent to the probabilities that no renewal occurs up to time $t$.

The initial conditions for the jump first model must be treated with some care.
As stated in Eq. (\ref{flux}) at time $t=0$ the system is prepared and the particle is at the origin.
We assume that a jump took place at $t=0^+$. In that sense the process begins at the start of the measurement.
Thus, if taking $t\to0^+$ one finds with Eq. (\ref{flux}) for the wait-first case Eq. (\ref{pdfw}) 
$p_{\text{w}}(x,0)=\delta(x)$ and for the jump-first case Eq. (\ref{pdfj}) $p_{\text{j}}(x,0)= f(x)$, as expected.

In particular, we will look at the linear coupling for the jump models, meaning that the waiting time is linearly proportional 
to the jump distance. This results in the simple coupling functions \cite{ZumKlaf93}
\begin{eqnarray}
f(x|\tau)=\delta(|x|-v_0\tau), \label{fcoup} \\
\psi(\tau|x)=\delta(\tau-v_0^{-1}|x|). \label{psicoup} 
\end{eqnarray}
We denote the proportionality constant by $v_0$ as it 
has the dimension of a velocity (and can indeed be considered as the speed of the single particle
if one measures the ratio of the distance travelled to the duration of the step).
The particular form of the primary waiting-time pdf $\psi(\tau)$ and jump-length pdf $f(x)$ 
will be specified later when needed. The following general considerations do not require such specification yet.

A different scenario is a velocity model (see Fig. \ref{RWmodels}), where during each step of duration $\tau$ drawn from $\psi(\tau)$ 
a particle moves with a constant speed $v$ drawn according to a velocity pdf $h(v)$. 
Equivalently to the loss flux in the jump model, we now write down the balance equation for the
probability density $\nu(x,t)$, which describes the frequency of velocity changes at point $x$ at a given time $t$
\begin{eqnarray}
 \nu(x,t) &=& \int_{-\infty}^{\infty}  dv \int_0^t \nu(x-v\tau,t-\tau)h(v)\psi(\tau) d\tau  
\nonumber\\
&&
+ \delta(x)\delta(t). \label{vchange}
\end{eqnarray}
A velocity change can occur in $x$ and at time $t$ if the previous velocity change to the value $v$
took place at $x-v\tau$ at the time $t-\tau$.
With this we get the particle density 
\begin{eqnarray}
p_{\text{v}}(x,t) &=&
\int_{-\infty}^{\infty}\! dv \int_0^t \nu(x-v\tau,t-\tau) h(v)\Psi(\tau) d\tau \label{pdfv}
\end{eqnarray}
where $\Psi(\tau)=1-\int_{0}^\tau \psi(\tau')d\tau'$ is again the probability not to have a renewal until $\tau$.

Note that in the special case of the velocity model (L\'evy walk) with just one speed $h(v)=1/2\left[\delta(v-v_0)+\delta(v+v_0)\right]$
\cite{Geisel85,Shle87,ZumKlaf93} the starting points and end points of the consecutive steps coincide with 
those of the jump models with linear coupling, although the paths in the $(x,t)$-plane are different (see Fig. \ref{RWmodels}). 
Therefore these models differ only by their last, incomplete step. 
As we proceed to show, this seemingly minor difference leads to dramatic effects on particles' pdf $p(x,t)$.
Beyond the fundamental interest in the long time limit of random walks, this issue is important also for computer generated
random walks, since in the normal situation, these definitions of sample paths all converge to unique behavior.


\section{Propagators}

\subsection{General expressions in Fourier-Laplace space}
Equations (\ref{flux}), (\ref{pdfw}) and (\ref{pdfj}) for the jump models and Eqs. (\ref{vchange}), (\ref{pdfv}) 
for the velocity model are solved with the help of Fourier-Laplace transforms \cite{SokKlafBook}. 
By setting the initial distribution of particles to the delta function $p(x,t=0)=\delta(x)$ we can find the propagators $G(x,t)$ 
for all three models. For the jump models, in Fourier-Laplace domain we get \cite{VZ2} (see Appendix A):
\begin{eqnarray}
G_{\text{j}}(k,s) &=& 
\frac{{\cal FL}\left\lbrace f(x)\Psi(\tau| x) \right\rbrace}{1-{\cal FL}\left\lbrace f(x)\psi(\tau|x)\right\rbrace};
\label{propfouj}\\
G_{\text{w}}(k,s) &=& 
\frac{{\cal L}\left\lbrace \Psi(\tau) \right\rbrace}{1-{\cal FL}\left\lbrace f(x|\tau)\psi(\tau)\right\rbrace};
\label{propfouw}
\end{eqnarray}
where $\cal{F}$ and $\cal{L}$ stand for Fourier- and Laplace transforms, with $k$ and $s$ denoting the coordinates 
conjugate to $x$ and $t$, respectively.

Using Eqs. (\ref{vchange}), (\ref{pdfv}) and convolution theorems for Fourier and Laplace transforms,
we get in accordance with previous studies \cite{KB,VZ1} the propagator 
of the velocity model $G_{\text{v}}$ which is given by
\begin{eqnarray}
G_{\text{v}}(k,s) &=& 
\frac{
{\cal L}
\left\lbrace \Psi(\tau) h(k\tau)\right\rbrace
}
{1 - {\cal L} \left\lbrace  \psi(\tau) h(k\tau) \right\rbrace} ,
\label{MW-mod}
\end{eqnarray}
where $k\tau$ is the Fourier variable conjugate to $v$ and hence the (spatial) Fourier transform of the velocity distribution 
${\cal F} \left\lbrace h(v)\right\rbrace = \int_{-\infty}^{\infty} e^{-ik\tau v} h(v) dv = h(k\tau)$.

Eq. (\ref{MW-mod}) retains the form of the well-known Montroll-Weiss \cite{MontWeiss} equation for the pdf of the 
(uncoupled) continuous time random walk (CTRW) to find the particle at $x$ at time $t$,
modified such that it applies to random jumps taking place not in space but in velocity \cite{footn1}.
Eq. (\ref{propfouw}) is a special case of an equation for CTRW introduced by Scher and Lax \cite{ScherLax73}, accounting for 
coupled waiting times and jump lengths. Eq. (\ref{propfouj}) represents a further modification of the latter, 
reflecting the fact that there the survival probability $\Psi(\tau|x)$ depends on the jump length.

The numerators in Eqs. (\ref{propfouj}-\ref{MW-mod}) reflect the effect of the last jump interval in the process
on the final position of the particle. This last incomplete time interval is called backward recurrence time $\tau_b = t-t_n$,
where $t_n$ is the epoch of the last renewal (see Fig. \ref{sketch}). 
As we will see later, this last interval has a crucial effect on the shape of the 
asymptotic particle pdfs, unlike usual random walk theories. Note that, though in a slightly different manner,
the statistics of the last jump event also plays an important role in the description of transport 
of cold atoms in optical lattices \cite{EliErez14}.

An analytical representation and a direct inversion of Eqs. (\ref{propfouj}-\ref{MW-mod}) is 
not feasible in most cases. As an alternative, one has to resort to 
the asymptotic analysis for large space and time scales $x,t\to\infty$. 
Going to Fourier/Laplace space using the Tauberian theorem \cite{MetzKlaf00}, 
this limit corresponds to $(k,s)\to(0,0)$, in our (ballistic) case
such that $(k/s)=\text{const.}$ However, even then the explicit analytical
Fourier - Laplace inversion is often not possible and has to be performed numerically.
In the following, we assume that the persistence time and jump length pdfs fall off like a power law:
\begin{equation}
\psi(\tau) = \frac{1}{\tau_0}\frac{\gamma}{(1+\tau/\tau_0)^{1+\gamma}},\quad 0<\gamma<1
\label{wtpdf}
\end{equation}
for the waiting time of the wait-first model and for the flight time of the velocity model, and
\begin{equation}
f(x) = \frac{1}{2x_0}\frac{\gamma}{(1+|x|/x_0)^{1+\gamma}},\quad 0<\gamma<1,
\label{jlpdf}
\end{equation}
for the jump length distribution in the jump-first model. 
Note that our final results are not sensitive to the specific small $\tau$ or -$x$ behavior. The power law tails 
completely determine the long time behavior which is rooted in the generalized central limit theorem, see for example
\cite{Bou90}. Thus our results are valid for the general class of waiting time or jump length pdfs that 
have the same asymptotic limit as Eq. (\ref{wtpdf}) or Eq. (\ref{jlpdf}), respectively.

Via the coupling relations (\ref{fcoup},\ref{psicoup}), the Eqs. (\ref{wtpdf},\ref{jlpdf}) 
fully determine also the step lengths of the wait-first, and time intervals of the jump-first model, respectively.
Thus in the wait-first case the full coupled jump pdf is 
\begin{equation}
K_{\text{w}}(x,\tau) = \psi(\tau)f(x|\tau) = 
\frac{1}{\tau_0}\frac{\gamma}{(1+\tau/\tau_0)^{1+\gamma}}\delta\left(|x|-v_0\tau\right)  , 
\end{equation}
from which we calculate the effective jump length pdf
\begin{equation}
f_{\text{w}}(|x|) = \int_0^\infty \Psi(x,\tau)d\tau = \frac{1}{v_0\tau_0}\frac{\gamma}{(1+|x|/(v_0\tau_0))^{1+\gamma}}
\end{equation}
which corresponds exactly to (\ref{jlpdf}) if we set $x_0=v_0\tau_0$.
An analogous derivation applies for the jump first model, so that indeed Eqs. (\ref{fcoup},\ref{psicoup}) 
together with (\ref{wtpdf},\ref{jlpdf}) yield the same effective jump length and waiting time distributions.
Thus $f_w(|x|)=\psi(\tau)|_{\tau=|x|}$ and $\psi_j(\tau) = f(|x|)|_{|x|=\tau}$ for $x_0=v_0\tau_0$ 
which allows us to compare between the two jump models. The construction of the simple L\'evy walk with 
two velocity states $h(v)=1/2[\delta(v-v_0)+\delta(v+v_0)]$ 
such that its steps all end in the same points in time and space as in the jump models 
requires to choose the same $v_0$ for all models. Later we will also see what changes in more complicated 
velocity models with distributed velocities.
The resultant (effective) waiting time pdfs of all these models lack a typical time scale since the mean waiting time 
diverges, and hence the overall motion of the particle will be governed by a few very large 
(of the order of the observation time) 
persistence time intervals during which the particle's state of motion does not change. 
Therefore this regime is referred to as the ballistic one \cite{Qdot2,DF13}.

\subsection{Fourier-Laplace inversion in the ballistic scaling regime}
When stating that the propagator of a random walk model has the ballistic scaling we imply 
that it can be written in the following form:
\begin{equation}
G(x,t) \cong \frac{1}{t} \Phi\left( \frac{x}{t}\right), \hspace{.5cm} t\to\infty \label{scaling}
\end{equation}
where $\Phi$ is the scaling function. In Fourier-Laplace space this would correspond to a similar relation
\begin{eqnarray}
G(k,s) &=& \frac{1}{s} g\left( \frac{ik}{s}\right). \label{invProp1}
\end{eqnarray}
For the Fourier-Laplace inversion of this expression we will follow a procedure 
similar to that given e.g. in \cite{GL01}.
By using the characteristic function and the definitions of the integral transforms 
we can write (see Appendix D)
\begin{eqnarray}
G(k,s) &=& 
\int_0^\infty e^{-st} 
\left\langle e^{-ikX}\right\rangle\;dt \nonumber\\
&=& \frac{1}{s}\left\langle \frac{1}{1 + \frac{ik}{s}Y}\right\rangle.
\label{GKS}
\end{eqnarray}
Here, angular brackets denote the averaging with respect to a random variable $X$ which has a pdf $P(X)$ 
\begin{equation}
\langle F(X)\rangle
=\int\limits_{-\infty}^{\infty}F(X)P(X)dX.
\end{equation}
In our case, $X$ is the coordinate of the particle at time $t$ and therefore $P(X)=G(X,t)$. 
In addition, we introduced a scaled variable $Y=X/t$ to obtain the second line of Eq.(\ref{GKS}). 
Now by comparing Eqs. (\ref{GKS}) and (\ref{invProp1}) we can identify the scaling function $g$ in Fourier-Laplace space as:
\begin{equation}
g(\xi) = \left\langle \frac{1}{1 + \xi Y}\right\rangle; \quad \xi\equiv\frac{ik}{s}. \label{genfc}
\end{equation}

%
%
%
%

%
We use the definition $\Phi(y)=\langle\delta(y-Y)\rangle$ and the Sokhotsky-Weierstrass theorem
\cite{footn2} to write
\begin{eqnarray}
\Phi(y) &=& \left\langle \delta(y-Y) \right\rangle = -\frac{1}{\pi} \lim_{\epsilon\to 0}
\mathrm{Im} \left\langle \frac{1}{y-Y+i\epsilon}\right\rangle \nonumber\\
&=&
-\frac{1}{\pi} \lim_{\epsilon\to 0}\mathrm{Im} \left[\frac{1}{y+i\epsilon} \left\langle \frac{1}{1-\frac{Y}{y+i\epsilon}}
\right\rangle\right] \nonumber
\end{eqnarray}
Finally, if we compare the above formula with Eq. (\ref{genfc}) we obtain the recipe to find the analytical 
expression of the scaling function $\Phi(y)$ if its counterpart in Fourier-Laplace space $g(\xi)$ is known:
\begin{eqnarray}
\Phi(y) &=& 
-\frac{1}{\pi} \lim_{\epsilon\to 0}\mathrm{Im} \left[\frac{1}{y+i\epsilon} g\left(-\frac{1}{y+i\epsilon}\right) \right] .
\label{Gsc}
\end{eqnarray}
A similar method was used before in \cite{GL01} for the inversion of a double Laplace transform. 
Here it is generalized to the case of time and two-sided space variables (as in \cite{Reb08}).
Below we demonstrate how it works in practice.

\section{Results for Jump Models}
\subsection{Two-sided jump models}
We start with general analytical expressions for the propagators of the jump models Eqs. (\ref{propfouj}) and (\ref{propfouw}), 
use the simple coupling relations Eqs. (\ref{fcoup},\ref{psicoup}) and proceed with the asymptotic analysis. 
For a waiting time distribution of the form as in Eq. (\ref{wtpdf}) in the long-time limit, its expansion in Laplace space 
is given by (and similarly for the jump length pdf Eq.(\ref{jlpdf}) in Fourier space)
\begin{equation}
\psi(s) \simeq 1 - \tau_0^\gamma\Gamma(1-\gamma)s^{\gamma} \label{wtdLap}
\end{equation}
After some algebra (see Appendix B) we find for the propagators in Fourier-Laplace domain 
in the small $k$ and $s$ limit
\begin{eqnarray}
G_{\text{j}}(k,s) &=&
\frac{1}{s} 
\left[
1 - \frac{\left(ikv_0\right)^{\gamma} + \left(-ikv_0\right)^{\gamma}}
{\left(s+ikv_0\right)^{\gamma} + \left(s - ikv_0\right)^{\gamma}}
\right]
\label{propj}
\end{eqnarray}
and
\begin{equation}
G_{\text{w}}(k,s) = \frac{1}{s}
\frac{2 s^{\gamma}}{\left[\left(s - ikv_0\right)^{\gamma} 
+ \left(s + ikv_0\right)^{\gamma}  \right]}.
\label{propw}
\end{equation}
We indeed see that these expressions assume the scaling form as in Eq. (\ref{invProp1}), from which we identify the scaling functions
\begin{equation}
g_{\text{j}}(\xi) = 1 - \frac{\left(-\xi\right)^{\gamma} + \left(\xi\right)^{\gamma}}{\left(1-\xi\right)^{\gamma} + \left(1+\xi\right)^{\gamma}}
\end{equation}
and
\begin{equation}
g_{\text{w}}(\xi) = \frac{2}{\left(1-\xi\right)^{\gamma} + \left(1+\xi\right)^{\gamma}}
\end{equation}
with $\xi=iv_0k/s$ so that the scaling variable becomes $y=x/(v_0t)$.
This allows us again to perform the Fourier-Laplace inversion in the scaling regime.
By Eqs.(\ref{Gsc}), (\ref{propj}) and (\ref{propw}) we find the scaling solutions in original space-time domain:
\begin{eqnarray}
&&\Phi_{\text{j}}(y) = \frac{\sin\left[\pi\gamma\right]}{\pi} 
y^{-1} \times \nonumber\\ 
&& 
\left\lbrace
\begin{array}{cc}
\frac{\mathrm{sign} y}{|y+1|^{\gamma} + |y-1|^{\gamma}}
&
1 \leq |y| < \infty 
\nonumber \\
\nonumber \\
\frac{|y+1|^{\gamma} - |y-1|^{\gamma}}{|y+1|^{2\gamma} + |y-1|^{2\gamma} + 2|y+1|^{\gamma}|y-1|^{\gamma}\cos\left[\pi\gamma\right]}
&
0 \leq |y| < 1 \nonumber
\end{array}
\right. \nonumber\\
&&\label{scj}
\end{eqnarray}
\begin{eqnarray}
&&\Phi_{\text{w}}(y) = \frac{2\sin\left[\pi\gamma\right]}{\pi} 
|y|^{\gamma-1} \times \nonumber\\ 
&& 
\left\lbrace
\begin{array}{cc}
0
&
1 \leq |y| < \infty 
\nonumber \\
\nonumber \\
\frac{\left(1-|y|\right)^{\gamma}}{|y+1|^{2\gamma} + |y-1|^{2\gamma} + 2|y+1|^{\gamma}|y-1|^{\gamma}\cos\left[\pi\gamma\right]}
&
0 \leq |y| < 1 \nonumber
\end{array}
\right. \nonumber\\
&&\label{scw}
\end{eqnarray}
where $y=x/(v_0t)$ is the scaling variable (see Fig. \ref{p2jw0}).
We also find from Eq. (\ref{propw}) a mean-squared displacement (MSD) 
$\left\langle x^2 \right\rangle_{\text{w}} = (1-\gamma)\gamma v_0^2 t^2/2$ or in terms of the 
scaling variable $\left\langle y^2 \right\rangle_{\text{w}} = (1-\gamma)\gamma/2$. Although the pdf for the jump-first 
model scales ballistically, it does not possess a finite MSD. 
This is to be expected, since the variance of the jump length
diverges and the jump is performed at the beginning of a waiting time - the last jump can be very large when the 
corresponding last waiting time is far from being completed.


\subsection{One-sided jump models}
\begin{figure}[h]
\centering{{\includegraphics[width=.30\textwidth]{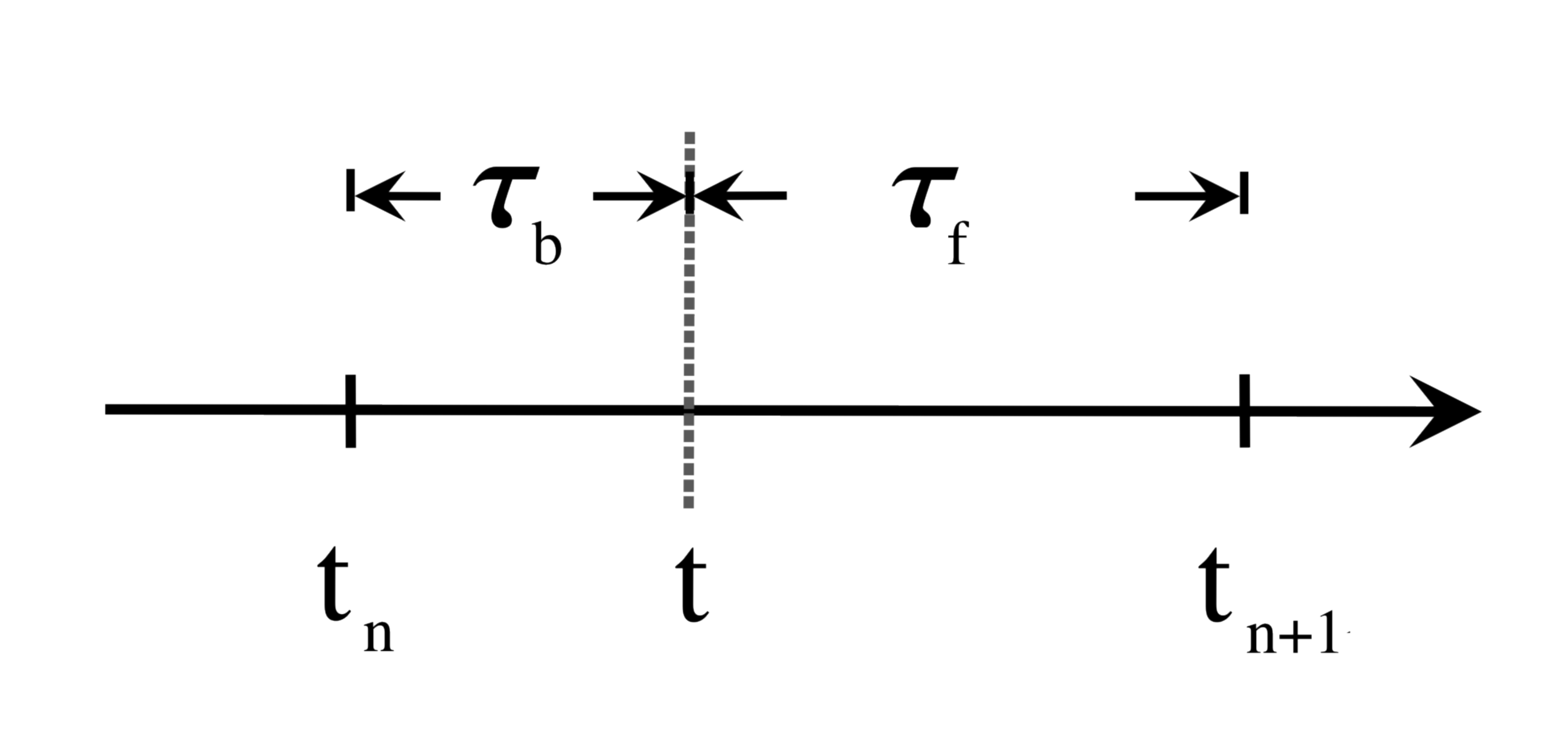}}
\caption{\label{sketch} Backward and forward recurrence times. Given a measurement time $t$,
the backward recurrence time $\tau_b$
is the time elapsed since the time $t_n$ the last event took place. The forward recurrence time $\tau_f$
is the time span between $t$ and the time $t_{n+1}$ at which the next renewal will take place.
}
}
\end{figure}

\begin{figure}[h]
\centering{{\includegraphics[width=.50\textwidth]{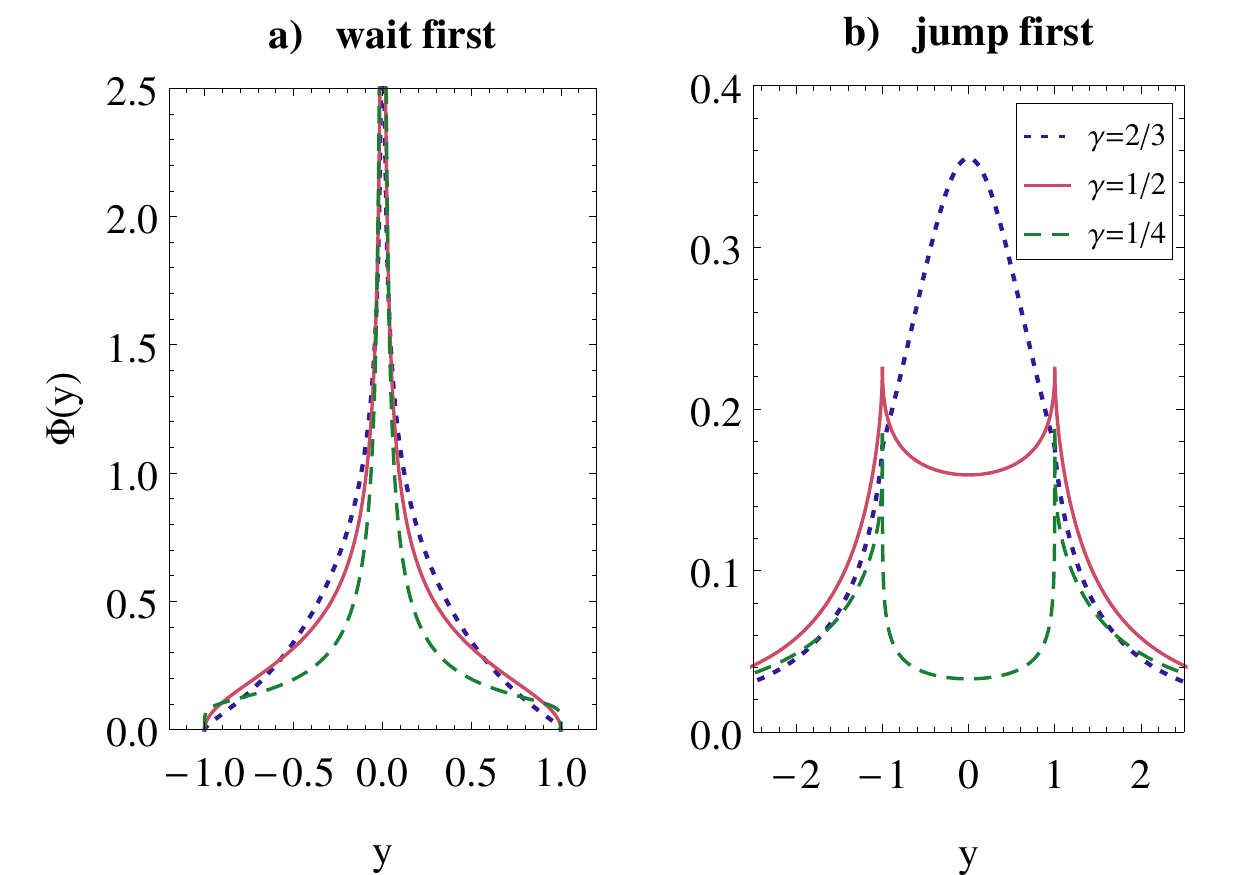}}
\caption{\label{p2jw0} (Color online) Propagators of two jump models in the ballistic scaling regime. 
The rescaled propagators for (a) the wait first and (b) the jump-first model from (Eqs. (\ref{scw}), (\ref{scj})) 
are shown for different values of the exponent $\gamma$ determining the power-law tail 
of the waiting time and jump length distributions. 
}
}
\end{figure}

\begin{figure}[h]
\centering{{\includegraphics[width=.50\textwidth]{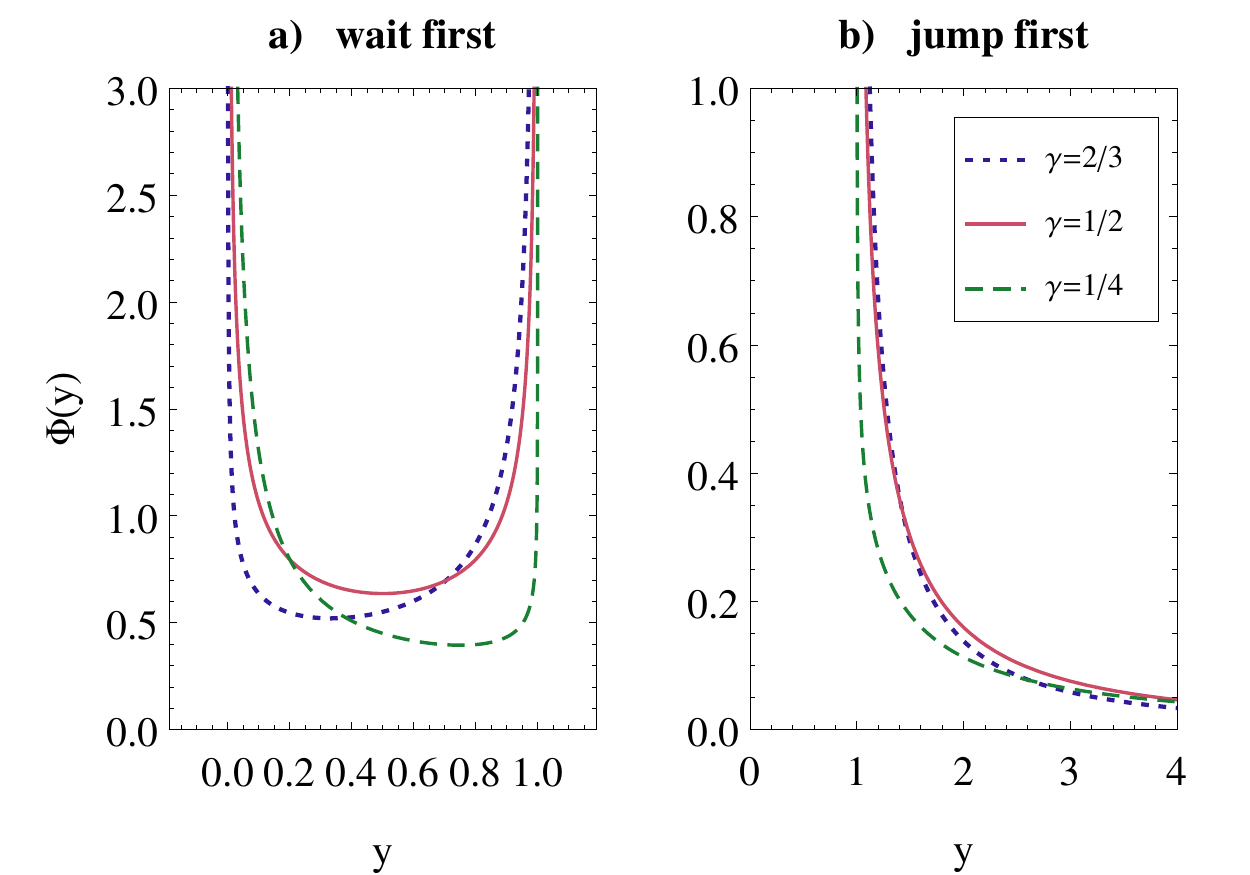}}
\caption{\label{p1jw0} (Color online) One-sided jump models. In this particular case only jumps to the 
right are allowed. One-sided wait-first (a) and jump-first model (b) (Eqs. (\ref{scwa}), (\ref{scja})), 
are shown for different values of the power-law tail exponent $\gamma$ governing the flight and jump length distributions. 
These results are related to the statistics of the  backward and forward recurrence times, as discussed in the text.}
}
\end{figure}

The reason for the remarkable shape differences of the scaled densities for the different models (see Fig. \ref{p2jw0} )
becomes immediately clear if we consider the following simplified example.
In what follows we allow for jumps only in the positive direction and indicate this in the scaling 
functions by a superscript ``+". 
The scaling functions become
\begin{eqnarray}
g_{\text{j}}^+(\xi) &=& 1- \frac{\xi^{\gamma}}{\left(1+\xi\right)^{\gamma}},
\nonumber\\
\nonumber \\
g_{\text{w}}^+(\xi) &=& \frac{1}{\left(1+\xi\right)^{\gamma}} .
\nonumber
\end{eqnarray}
Thus, the one-sided scaling solutions are
\begin{eqnarray}
&&\Phi_{\text{j}}^+(y) = 
\left\lbrace
\begin{array}{cc}
\frac{\sin\left[\pi\gamma\right]}{\pi} y^{-1}\left(y-1\right)^{-\gamma}
&
1 < y < \infty 
\nonumber \\
\nonumber \\
0
& 
-\infty \leq y \leq 1 
\nonumber
\end{array}
\right. \nonumber \\
&&\label{scja}
\end{eqnarray}
\begin{eqnarray}
&&\Phi_{\text{w}}^+(y) = 
\left\lbrace
\begin{array}{cc}
0
&
1 < y < \infty 
\nonumber \\
\nonumber \\
\frac{\sin\left[\pi\gamma\right]}{\pi} \left(1-y\right)^{-\gamma} y^{\gamma-1}
& 
-\infty \leq y \leq 1 \nonumber
\end{array}
\right.\nonumber\\
&&\label{scwa}
\end{eqnarray}
as shown in Fig. \ref{p1jw0}. The MSD of the jump-first model again diverges. For the one-sided wait-first
model $\left\langle x^2 \right\rangle_{\text{w}} = (1+\gamma)\gamma v_0^2 t^2/2$ and thus 
$\left\langle y^2 \right\rangle_{\text{w}} = (1+\gamma)\gamma/2$, or $\text{var}(y)_{\text{w}}=(1-\gamma)\gamma/2$,
as $\left\langle y \right\rangle_{\text{w}} = \gamma$.
The propagators of the one-sided jump models are closely related to the distributions of forward and
backward recurrence times $\tau_f$ and $\tau_b$, i.e. the time interval from the measurement time to the next renewal and the
time interval that passed since the last renewal, respectively (see Fig. \ref{sketch}). 
The positions $x$ in the jump and wait first models are
$x^{(\text{j})} = v_0(t + \tau_f)$, $x^{(\text{w})} = v_0(t - \tau_b)$
and thus
\begin{eqnarray}
y^{(\text{j})} = 1 + \frac{\tau_f}{t}, \label{tauf}\\
y^{(\text{w})} = 1 - \frac{\tau_b}{t}. \label{taub}
\end{eqnarray}
Indeed, the forward and backward recurrence time scaling distributions of the variables $y^{(f)}=\tau_f/t$
and $y^{(b)}=\tau_b/t$ are obtained 
by performing the above transformations Eqs. (\ref{tauf}), (\ref{taub})
of the scaling variables $y$ in Eqs. (\ref{scja}), (\ref{scwa}), respectively (see also \cite{Feller,GL01}). 
This example demonstrates strikingly that the difference between the models 
comes into effect only through the last renewal period.

%

%
%

\section{Results for the Velocity Models}

The propagator for the velocity model Eq. (\ref{MW-mod}) can be rewritten as
\begin{eqnarray}
G_{\text{v}}(k,s) &=& \frac{\int_{-\infty}^\infty dv \Psi(s+ikv) h(v)}{1 - \int_{-\infty}^\infty dv \psi(s+ikv) h(v)},
\label{Prop}
\end{eqnarray}
where the $\Psi(s+ikv)$ and $\psi(s+ikv)$ are Laplace-transforms and the shift theorem was used (see Appendix C).
This result is exact, provided both integrals in Eq.(\ref{Prop}) converge.
In the ballistic regime $0<\gamma<1$ we use the expansion given in Eq. (\ref{wtdLap}) to obtain 
the asymptotic version of this result
\begin{equation}
G_{\text{v}}(k,s)=
\frac{1}{s} 
\frac{\int_{-\infty}^\infty \left(1+ikv/s\right)^{\gamma-1} h(v) dv}
{\int_{-\infty}^\infty \left(1+ikv/s\right)^{\gamma} h(v) dv} . \label{invpropApp}
\end{equation}
By comparing it with the scaling form of Eq. (\ref{invProp1}) we obtain
\begin{eqnarray}
\nonumber \\
g_{\text{v}}\left(\xi\right) &=& \frac{\int_{-\infty}^\infty \left(1+\xi v\right)^{\gamma-1} h(v) dv}
{\int_{-\infty}^\infty \left(1+\xi v\right)^{\gamma} h(v) dv} .
\end{eqnarray}
Finally, performing the inversion in the scaling regime according to Eq. (\ref{Gsc}) yields
\begin{eqnarray}
\Phi_{\text{v}}(y)&=& -\frac{1}{\pi}\lim_{\epsilon\to 0} \mathrm{Im}
\frac{\int_{-\infty}^\infty\left(y+i\epsilon-v\right)^{\gamma-1} h(v) dv }
{\int_{-\infty}^\infty  \left(y+i\epsilon-v\right)^{\gamma} h(v) dv} . \label{Gscint}
\end{eqnarray}

We should note that the problem of finding the propagator in the velocity model 
corresponds to the problem of time-averaging the position of a subdiffusing particle subject to 
an external binding potential as investigated in detail in \cite{Reb08}: 
Indeed, we can view the velocity model as a decoupled CTRW in velocity
space, and the inference of the particle position requires an integration over time. 
The scaled (here $v_0=1$) position $y=x/t$ of the single particle after a time $t$, provided it started at 
$t_0=0$, $x_0=0$ is given by 
\begin{eqnarray}
y(t) = \frac{1}{t}\int_0^t v(t^\prime) dt^\prime .
\end{eqnarray}
Thus, for the time averaged CTRW, $h(v)$ corresponds to the Boltzmann equilibrium 
or steady state distribution in space \cite{Reb08}.

%

\begin{figure}[h]
\centering{
{\includegraphics[width=.44\textwidth]{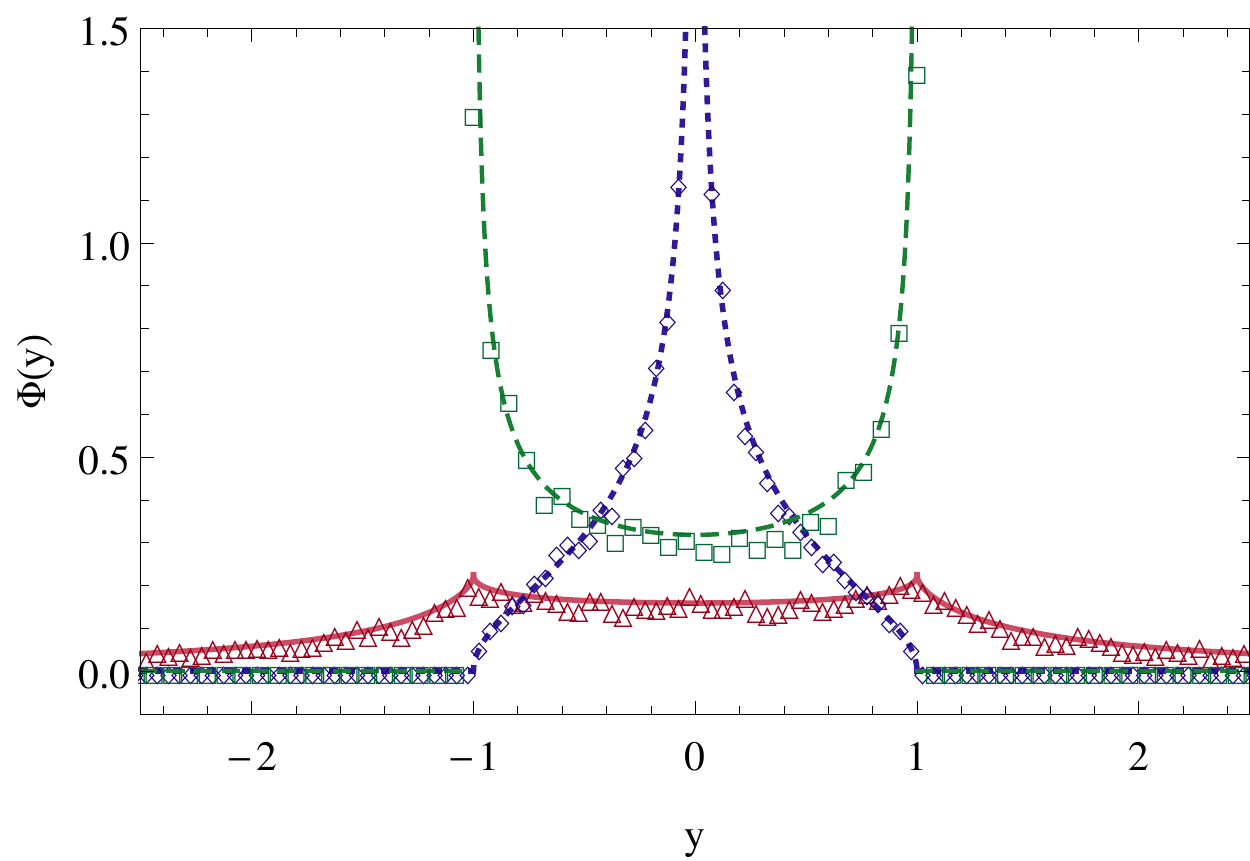}}
\caption{\label{N} (color online) Theory and simulations for the wait-first model Eq. (\ref{scw}) (solid line, triangles, red),
the jump first model Eq. (\ref{scj}) (dotted line, diamonds, blue) and 
the two-state velocity model Eq. (\ref{lamperti}) (dashed line, squares, green). 
$\gamma=1/2$, scaling variable $y=x/(v_0t)$.}
}
\end{figure}

%
Let us now furnish the above result with some examples. 
We start with the two-state velocity pdf $h(v) = \left[\delta(v-v_0) + \delta(v+v_0) \right]/2$. 
By repeating the same sequence of steps as for the jump models we arrive at
%
\begin{eqnarray}
g_{\text{v}}(\xi) &=& \frac{\left( 1-\xi  \right)^{\gamma-1} + \left( 1+\xi  \right)^{\gamma-1} }
{\left( 1-\xi  \right)^{\gamma} + \left( 1+\xi  \right)^{\gamma}},
\end{eqnarray}
$\xi=ikv_0/s$.
With Eq. (\ref{Gsc}) the scaling form of the propagator is found to be the
Lamperti-distribution \cite{Lamperti}
\begin{eqnarray}
&&\Phi_{\text{v}}(y) =\frac{\sin\pi\gamma}{\pi} \times\nonumber \\ 
&&
\frac{|y-1|^{\gamma}|y+1|^{\gamma-1} + |y+1|^\gamma|y-1|^{\gamma-1}}
{|y-1|^{2\gamma} + |y+1|^{2\gamma} + 2|y-1|^\gamma|y+1|^\gamma \cos\pi\gamma}
\label{lamperti}
\end{eqnarray}
where $y=x/(v_0t)$ and $|y|<1$, and $\left\langle y^2 \right\rangle =(1-\gamma)$. 
This distribution plays a role, for example, in the prediction of the
time averaged intensity of the light emitted by a blinking quantum dot \cite{Qdot}.
For $\gamma=1/2$ this scaling distribution assumes a particularly simple form,
\begin{eqnarray}
&&\Phi_{\text{v}}(y) = \frac{1}{\pi \left( 1-y^2 \right)^{1/2}}\,,\hspace{.4cm} |y|<1  ,\label{delta}
\end{eqnarray}
the well known arcsine distribution \cite{Feller}.
It is instructive to compare the propagator of this type of L\'{e}vy walk with those of the two jump models. 
As it is clear from Fig. 1, the trajectory of the walk passes through the same end points as both jump models, 
yet it does it by a different path in the $(x,t)$ plane. As a result, the shape of the propagator is very distinct, 
see Fig. \ref{N}. As before, the origin of the difference is in the last unfinished step: 
in case of the corresponding one sided walk, i.e. the walk with just one velocity $h(v)+\delta(v-v_0)$,
the position of the particle is given by a simple $x^{(v)}=v_0t$, different from both jump models. 
We also use Fig. \ref{N} to compare our analytical results with direct numerical simulations of random walk models 
which show excellent agreement. Fig. \ref{N} clearly demonstrates that the particles spread further in the jump
first model if compared with the velocity- and wait first model, the latter being the slowest process.
%

%
%

The distinct feature of the velocity model is the ability to include the velocity distribution of the random walking particles. 
Below, we show how different velocity pdfs change the scaling shape of the corresponding propagator. 
For illustration, in addition to the two state velocity pdf, we chose a uniform velocity distribution on a finite interval, 
Gaussian, and Cauchy distributions.

\begin{figure}[h]
\centering{
{\includegraphics[width=.46\textwidth]{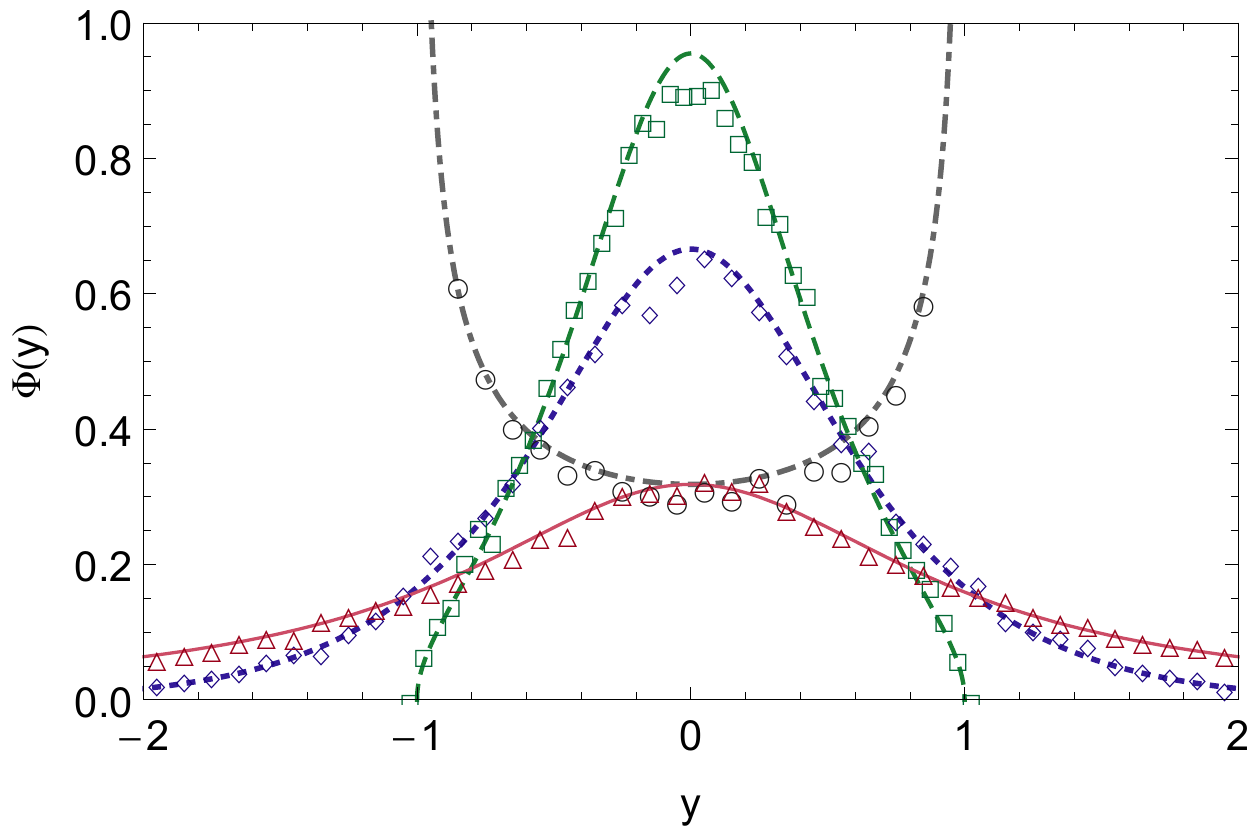}}
\caption{\label{velox} (Color online) Scaling functions of the random walks with random velocities model.
Results for four different velocity pdfs of the walking particles are depicted:
two state velocity $v=\pm v_0$ Eq. (\ref{delta}) (dash-dot, circles, grey), 
uniform on an interval Eq. (\ref{box}) $[-v_0,v_0]$ (dashed, squares, green), 
Gaussian Eq. (\ref{Gscint}) (dotted, diamonds, blue) and 
Cauchy Eq. (\ref{cauchy1}) (full line, triangles, red); $\gamma=1/2$. 
The result for the Gaussian velocity pdf was obtained using {\sc Mathematica}. 
Theory and simulations nicely match without fitting.
}
}
\end{figure}

For the uniform velocity pdf between $\pm v_0$, 
${\theta(v_0-|v|)/(2v_0)}$ where $\theta$ is the 
Heaviside step function, we find with $\xi = v_0 ik/s$
\begin{eqnarray}
g_{\text{v}}(\xi) &=& \frac{(1+\gamma)}{\gamma}
\frac{\left( \left( 1-\xi\right)^{\gamma}   - \left( 1+\xi\right)^{1\gamma} \right) }
{\left( 1-\xi\right)^{1+\gamma}   - \left( 1+\xi\right)^{1+\gamma} }
\end{eqnarray}
and
\begin{eqnarray}
&&\Phi_{\text{v}}(y) = \frac{2(1+\gamma)\sin\pi\gamma}{\pi\gamma} \times \nonumber\\
&&\frac{\left(1-y^2\right)^\gamma}
{\left(1-y\right)^{2+2\gamma} + \left(1+y\right)^{2+2\gamma} + 2\left(1-y^2\right)^{1+\gamma}\cos\pi\gamma},
\label{box0}
\end{eqnarray}
with the MSD $\left\langle y^2 \right\rangle = 1/3(1-\gamma)$.
For $\gamma=1/2$ it reduces to a very simple expression
\begin{eqnarray}
&&\Phi_{\text{v}}(y) = \frac{3\sqrt{1-y^2}}{\pi\left(1+3y^2\right)}
\label{box}
\end{eqnarray}
shown in Fig. \ref{velox} (dashed, squares, green). 
As we see, similarly to the L\'{e}vy walk case with two velocity states, the propagator is bounded by ballistic 
fronts corresponding to the maximal possible speed. However, it has more of a bell shape profile, 
unlike the shape of the single-speed L\'evy walk
with its distinct infinite peaks at $|y|=1$.

%
%

For a Gaussian velocity pdf 
${h(v) = (2\pi v_0^2)^{-1/2}\exp\left[-v^2/(2v_0^2) \right]}$ we get a more involved expression
\begin{eqnarray}
&&g_{\text{v}}(\xi) = \nonumber\\
&&\frac{\int_{-\infty}^\infty \left[1+\xi v \right] ^{-1/2}  
\exp\left[-\frac{v^2}{2v_0^2}\right]  dv}{\int_{-\infty}^\infty 
\left[1+\xi v \right] ^{1/2} \exp\left[-\frac{v^2}{2v_0^2}\right]  dv}.
\end{eqnarray}
For $\gamma=1/2$, the corresponding Eq. (\ref{Gscint}) can be solved (for example in {\sc Mathematica}) and expressed in terms 
of hypergeometric (or Kummer's-) functions of the first and second kind, see Fig. \ref{velox} (dotted, diamonds, blue). 
Since the Gaussian pdf, in principle, allows for infinite speeds, the profile of the propagator this time is unbounded.
Still the MSD remains finite, $\left\langle y^2 \right\rangle = 1-\gamma$. 
Indeed, if the second moment of a symmetric $h_{sc}(v)=h(v/v_0)/v_0$ exists, 
we always have $\left\langle y^2 \right\rangle = (1-\gamma)\left\langle v^2 \right\rangle/v_0^2$, 
see also \cite{Reb08}.

%
%

A very special situation is induced by Cauchy-distributed velocities, $h(v)=1/(\pi v_0(1+(v/v_0)^2))$. 
It was shown in Ref. \cite{VZ1} that in this case the flight time distribution, in particular the value of $\gamma$,
has no effect on the resulting propagator, 
which in turn is also Cauchy. 
To show this, let us consider the exact answer for the velocity model as in Eq.(\ref{MW-mod}) and substitute 
the Fourier transform of the Cauchy velocity distribution: 
${h(k\tau)=\exp(-v_0|k|\tau)}$. After simple algebra, and without prescribing the particular form 
of $\psi(\tau)$ we arrive at:
\begin{equation}
G_{\text{v}}(k,s)=\frac{1}{s+v_0|k|}\label{cauchy_fl}.
\end{equation}
One can immediately recognize that the inverse Laplace and Fourier transform of Eq.(\ref{cauchy_fl}) will again 
lead to the Cauchy distribution:
\begin{equation}
G_{\text{v}}(x,t)=\frac{v_0t}{\pi(v_0^2t^2+x^2)}.\label{cauchy1}
\end{equation}
Therefore the scaling function, which we plot on Fig. \ref{velox} (full line, triangles, red), has the simple form 
$\Phi_{\text{v}}(y)=[\pi(1+y^2)]^{-1}$, and its MSD clearly diverges.
For such a simple answer for the propagator in Fourier-Laplace space, Eq.(\ref{cauchy_fl}), the application of the 
proposed inversion method becomes somewhat redundant, however, it can be demonstrated that it works here as well. 
To conclude this section we note that the model of random walks is very sensitive to the shape of the velocity distribution 
which gets reflected in the profile of the asymptotic density. This was noticed before \cite{Reb08}, as well as a 
similar phenomenon in the sub-ballistic, superdiffusive regime \cite{Reb14ICD}. 
However, in the subballistic case the effect was much weaker as it appeared only at the far tails of the distribution, 
whereas in the ballistic regime it dominates the whole propagator.

%


\section{Discussion}

We considered the long time scaling limit of the density profiles of particles for a large class of
one dimensional random walks that operate in the ballistic regime, which implies a scaling variable $y\sim x/t$.
Depending on the model, these densities differ strikingly. It is important to note that especially the 
last renewal period plays a crucial role with regard to these differences.
If we compare the jump models and the simple velocity model with constant 
speed where at the renewal only the direction of motion is chosen, we find that
exactly at the instant of a renewal the jump-first, wait-first and the simple velocity model 
are the same (Figure \ref{RWmodels}).
Thus they differ only by their last renewal interval, 
and this is the origin of the difference between their densities.

For both the one and two sided wait-first models the propagator is 
restricted to a finite interval of the scaling variable, $y\in(0,1)$ and $y\in[-1,1]$, 
respectively, since the particles can never overcome the front $|x|=v_0t$.
In contrast to that, in the one sided jump-first model the particles jump further ahead whenever the front
catches up with them, therefore $y\in(1,\infty)$. In the two-sided jump-first model we have 
$y\in(-\infty,\infty)$.
The propagators of both jump-first models exhibit heavy tails (Figs. \ref{p1jw0} and \ref{p2jw0}, right panels)
which renders them somewhat unphysical. Nevertheless they can serve as an impressive demonstration of the 
large effect of the final jump.

Whether the scaled propagators of the velocity model have a heavy tail or live on a finite domain
depends directly on the underlying velocity pdfs which can exhibit 
heavy tails or are constricted to a finite interval.
We explicitly calculated the scaled densities for some velocity- and jump-models, for which we
found excellent agreement with the results of direct Monte Carlo simulations of the respective processes. 
Although we considered $\delta$-function-like coupling between the jump distance and time it takes, 
the method that we suggested only requires the existence of the ballistic scaling. 
Therefore it can in principle be applied to other, distributed couplings as long as they lead to the ballistic regime. 
The mathematical machinery behind the method is specific to the ballistic regime, and it will be necessary to
develop other approaches for different scaling regimes.
We believe that the analytical results presented here provide an important step in our 
quantitative understanding of random walks and will facilitate the 
implementation of these random walk models in physics and other interdisciplinary applications.

\begin{acknowledgements}
M. S. was supported by the DFG within the Emmy Noether program (Schm2657/2). 
E. B. thanks the Israel Science Foundation
and the Max Planck Institute for the Physics of Complex Systems.
\end{acknowledgements}

\appendix

\section{Jump Models: Propagator in Fourier-Laplace Domain}

Fourier-Laplace transformation of Eq. (\ref{flux}) yields
\begin{eqnarray}
Q_\alpha(k,s) &=& K_\alpha (k,s)Q_\alpha(k,s) + 1 \nonumber\\
Q_\alpha(k,s) &=& \frac{1}{1-K_\alpha (k,s)}
\label{fluxFL}
\end{eqnarray}
where $K_{\text{j}}(k,s) = {\cal F}{\cal L}\left\lbrace f(x)\psi(\tau|x) \right\rbrace$ and
$K_{\text{w}}(k,s) = {\cal F}{\cal L}\left\lbrace f(x|\tau)\psi(\tau) \right\rbrace$.
The Fourier-Laplace transform is defined as 
${\cal F}{\cal L}\left\lbrace\cdot\right\rbrace = \int_{-\infty}^{\infty} dx \int_0^\infty dt e^{-ikx}e^{-st}(\cdot)$.
For Eqs. (\ref{pdfw}), (\ref{pdfj}) we have
\begin{eqnarray}
p_{\text{w}}(k,s) &=& {\cal L}\left\lbrace\Psi(\tau)\right\rbrace Q_w(k,s) \\
p_{\text{j}}(k,s) &=& {\cal F L}\left\lbrace f(x)\Psi(\tau|x)\right\rbrace Q_w(k,s)
\end{eqnarray}
Inserting (\ref{fluxFL}) results in Eqs. (\ref{propfouj}) and (\ref{propfouw}).

\section{Jump Models: Formal scaling solution}

Eqs. (\ref{propfouj}), (\ref{propfouw}) simplify due to the shift theorem for linear coupling, Eqs. (\ref{fcoup}), (\ref{psicoup}): 
\begin{eqnarray}
G(k,s)&=& \frac{F(k,s)}
{1-{\cal F}\left\lbrace \exp\left[-s\frac{|x|}{v}\right] f\left(|x|\right)\right\rbrace }, \label{propFouLa}
\end{eqnarray}
where $f$ is the coupling function between jump length and waiting time.
For the jump-first model,
\begin{eqnarray}
F_j(k,s) &=& {\cal F}\left\lbrace \frac{f(|x|)}{s} \left( 1-\exp\left[-s\frac{|x|}{v} \right] \right) \right\rbrace
\label{numj}
\end{eqnarray}
and
\begin{eqnarray}
F_w(k,s) &=& \frac{1-\psi(s)}{s}
\label{numw}
\end{eqnarray}
for the wait-first model.
With Eq. (\ref{wtdLap}) the denominator in (\ref{propFouLa}) becomes
\begin{eqnarray}
 1 &-& \int_{-\infty}^0 e^{\frac{sx}{v_0}} \frac{1}{2v_0} \psi\left(-\frac{x}{v_0}\right)e^{-\mathrm{i}kx} \; dx
\nonumber\\
&-& \int_{0}^{\infty} e^{-\frac{sx}{v_0}} \frac{1}{2v_0} \psi\left(\frac{x}{v_0}\right)e^{-\mathrm{i}kx} \; dx 
\nonumber\\
\simeq
&& \frac{\tau_0^\gamma\Gamma(1-\gamma)}{2} \left[\left(s+\mathrm{i}kv_0\right)^{\gamma} + \left(s - \mathrm{i}kv_0\right)^{\gamma}\right]
\label{den}
\end{eqnarray}
Correspondingly, we have for Eq. (\ref{numj})
\begin{eqnarray}
 && \frac{1}{s} 
\left[ 
\int_{-\infty}^0 \frac{1}{2v_0} 
\psi\left(-\frac{x}{v_0}\right)\left(1-e^{\frac{sx}{v_0}}\right) 
e^{-\mathrm{i}kx} \; dx \right. \nonumber\\
&& + \left.
\int_{0}^{\infty} \psi\left(\frac{x}{v_0}\right)\left(1-e^{-\frac{sx}{v_0}}\right) e^{-\mathrm{i}kx} \; dx 
 \right]
 \nonumber\\
&\simeq& \frac{1}{s}
\left[ 1 - \frac{\tau_0^\gamma\Gamma(1-\gamma)}{2} 
\left[\left(+\mathrm{i}kv_0\right)^{\gamma} + \left(- \mathrm{i}kv_0\right)^{\gamma}\right]\right.
\nonumber\\
&& - \left. 1 + \frac{\tau_0^\gamma\Gamma(1-\gamma)}{2} 
\left[\left(s+\mathrm{i}kv_0\right)^{\gamma} + \left(s - \mathrm{i}kv_0\right)^{\gamma}\right]
\right]
\label{numj1}
\end{eqnarray}
and for Eq. (\ref{numw})
\begin{equation}
\frac{1 - \psi(s)}{s} \simeq \frac{1}{s}\tau_0^\gamma\Gamma(1-\gamma)s^{\gamma}
\label{numw1}
\end{equation}
Inserting this back into Eq. (\ref{propFouLa}) yields Eqs. (\ref{propj}) and (\ref{propw}), respectively.

\section{Propagator in the Velocity Model}

Let us write explicitly the numerator in the fraction of Eq. (\ref{MW-mod})
\begin{eqnarray}
&&{\cal L}
\left\lbrace \Psi(\tau) h(k\tau)\right\rbrace (k,s) \nonumber \\
&&=
\int_0^\infty d\tau \int_{-\infty}^\infty dv e^{-i k\tau v} e^{-s\tau} \Psi(\tau) h(v) \nonumber \\
&&=
\int_{-\infty}^\infty dv \left( \int_0^\infty d\tau e^{-(s+i k v)\tau} \Psi(\tau) h(v)\right)  \nonumber \\
&&= \int_{-\infty}^\infty dv \Psi(s+ikv) h(v).
\label{MWnum}
\end{eqnarray}
Analogously for the denominator
\begin{eqnarray}
&&1 - {\cal L}
\left\lbrace \psi(\tau) h(k\tau)\right\rbrace (k,s) \nonumber \\
&&= 1 - \int_{-\infty}^\infty dv \psi(s+ikv) h(v).
\label{MWden}
\end{eqnarray}
Inserting this into Eq. (\ref{MW-mod}) yields Eq. (\ref{Prop}).

\section{Fourier-Laplace transform of the scaling function}

Explicitly, the Fourier-Laplace transform of Eq. (\ref{scaling}) is 
\begin{equation}
\int_{-\infty}^{\infty} \int_0^{\infty} \exp\left[-ikx-st\right] \frac{1}{t} \Phi\left(\frac{x}{t}\right) \,dt \,dx
\end{equation}
Subsequent variable transformation leads to
\begin{eqnarray}
&&\int_{-\infty}^{\infty} \int_0^{\infty} \exp\left[-(iky+s)t \right] \Phi\left(y\right) \,dt \,dy \nonumber\\
&&= \int_{-\infty}^{\infty}\frac{\Phi\left(y\right)}{iky+s} \,dy 
= \frac{1}{s}\int_{-\infty}^{\infty}\frac{\Phi\left(y\right)}{\frac{ik}{s}y+1} \,dy
\end{eqnarray}
which is equivalent to Eq. (\ref{GKS}).


\begin{thebibliography}{99}
\bibitem{footn1} For comparison write $k\tau = k^*$, then $n(k^*,s) = 
\left({\cal L} \left\lbrace \Psi(\tau) h(k^*)\right\rbrace (k,s)\right)/
\left(1 - {\cal} L \left\lbrace  \psi(\tau) h(k^*) \right\rbrace (k,s)\right) =
\left(\left(1-\psi(s)\right)/s\right) \left(1/\left(1-\psi(s) h(k^*)\right)\right) n(t=0,k^*)$
where the initial condition is modified to account for the velocity pdf, 
$ n(t=0,k^*,k) = h(k^*) {\cal F}_k\left\lbrace n(t=0,k)\right\rbrace $
\end{thebibliography}
\end{document}